\documentclass[11pt,a4paper]{article}
\pdfoutput=1
\usepackage{jcappub}
\usepackage{cprotect}

\newcommand{\PRE}[1]{{#1}} 

\newcommand{\be}{\begin{equation}}
\newcommand{\ee}{\end{equation}}
\newcommand{\bea}{\begin{eqnarray}}
\newcommand{\eea}{\end{eqnarray}}

\def\gev{\text{ GeV}}
\def\mev{\text{ MeV}}

\def\pb{\text{ pb}}

\def\kT{\text{ kT}}
\def\sr{\text{ sr}}
\def\cm{\text{ cm}}
\def\m{\text{ m}}

\def\s{\text{ s}}
\def\yr{\text{ yr}}

\newcommand{\Ar}[1]{{}^{#1}\!\text{Ar}}
\def\beq{\begin{eqnarray}}
\def\eeq{\end{eqnarray}}
\def\bea{\begin{eqnarray}}
\def\eea{\end{eqnarray}}

\def\sigmaSD{\sigma_{\rm SD}}

\newcommand{\gsim}{\lower.7ex\hbox{$\;\stackrel{\textstyle>}{\sim}\;$}}
\newcommand{\lsim}{\lower.7ex\hbox{$\;\stackrel{\textstyle<}{\sim}\;$}}

\usepackage{enumerate}
\include{epsf}

\title{
\textsc{Directional Searches at DUNE for Sub-GeV Monoenergetic Neutrinos Arising from Dark Matter Annihilation in the Sun}
\PRE{\vspace*{0.1in}}
}

\author[a]{Carsten Rott}
\author[a]{Seongjin In}
\author[b]{Jason Kumar}
\author[c]{David Yaylali}

\affiliation[a]{\mbox{Department of Physics, Sungkyunkwan University,} \\ \mbox{2066 Seobu-ro, Suwon 440-746, Korea}}
\affiliation[b]{\mbox{Department of Physics \& Astronomy, University of
Hawai'i,} \\ \mbox{2505 Correa Road, Honolulu, HI 96822, U.S.A.}}
\affiliation[c]{\mbox{Department of Physics, University of Arizona,} \\ \mbox{1118 E. Fourth Street, Tucson, AZ 85721, U.S.A.}
\PRE{\vspace*{.1in}}
}

\emailAdd{rott@skku.edu}
\emailAdd{seongjin.in@gmail.com}
\emailAdd{jkumar@hawaii.edu}
\emailAdd{yaylali@email.arizona.edu}

\abstract{
We consider the use of directionality in the search for monoenergetic sub-GeV neutrinos arising
from the decay of stopped kaons, which can be produced by dark matter annihilation in the core of
the Sun.  When these neutrinos undergo charged-current interactions with a nucleus at a neutrino
detector, they often eject a proton which is highly peaked in the forward direction.  The direction
of this track can be measured at DUNE, allowing one to distinguish signal from background by comparing
on-source and off-source event rates.  We find that directional information can enhance the signal to background ratio by up to a factor of 5.
}


\begin{document}

\begin{flushright}
{\large \tt
UH-511-1264-16  CETUP2016-004}
\end{flushright}

\maketitle
\flushbottom
\date{\today}


\section{Introduction}

One of the major strategies used to study interactions of dark matter
with Standard Model matter is the search for neutrinos arising from
dark matter annihilation in the core of the Sun~\cite{Silk:1985ax,Press:1985ug,Krauss:1985ks}.
The usual focus of this strategy is on
the production of energetic neutrinos from the decay of certain high-mass annihilation products, which is the subject of all present
experimental searches~\cite{Aartsen:2012kia,Choi:2015ara,Adrian-Martinez:2013ayv}. The high-energy
neutrino signal will almost certainly be accompanied by a low-energy neutrino component, which has its origin
in a hadronic cascade that develops in the dense solar medium and
produces large numbers of light long-lived mesons --- predominantly $\pi^+$ and $K^+$ --- which
will eventually stop and decay at rest.
The high-energy signal benefits from a large interaction cross section at the detector and faces smaller backgrounds compared
to less energetic neutrinos from the hadronic cascades.
However, if dark matter
annihilates predominantly to light quarks ($u$, $d$, and $s$) the low-energy neutrino signal might be much more
detectable. Although these channels produce few energetic neutrinos, the decays
of long-lived stopped mesons will produce
a large number of low-energy neutrinos, which can
be detected above background~\cite{Rott:2012qb,Bernal:2012qh}.  Moreover,
the decay of each $\pi^+$ or $K^+$ will produce a monoenergetic $\nu_\mu$ with an
energy $30\mev$ or $236\mev$, respectively.  These monoenergetic neutrinos would be a
striking signal at neutrino detectors with excellent energy resolution~\cite{Kumar:2015nja,Rott:2015nma}.
Liquid argon based detectors such as DUNE~\cite{Acciarri:2015uup} are ideally suited for this
search, but interesting prospects are also expected for liquid scintillator
based detectors such as RENO-50~\cite{Kim:2014rfa} or JUNO~\cite{Djurcic:2015vqa},
water-based liquid scintillators~\cite{Alonso:2014fwf}, or water cherenkov detectors such as
Hyper-Kamiokande (Hyper-K)~\cite{Abe:2015zbg,Abe:2011ts}.

When searching for high-energy neutrinos produced by dark matter annihilation, one
typically reduces the roughly isotropic atmospheric neutrino background by focusing
only on events wherein a charged-current
interaction in the detector produces a charged lepton pointing away from the Sun.
But this strategy fails if the neutrinos have energy $\lesssim {\cal O}(1)\gev$, because in
this case the charged leptons are produced largely isotropically.  In this work, we
point out that directional information for sub-GeV neutrinos can be determined from
the recoil of the struck nucleon within the nuclear target, which is typically liberated
from the nucleus in the forward direction.
The use of this directional information to remove background can enhance the sensitivity of this dark matter search strategy.  More importantly, it significantly enhances signal to background; it allows one to estimate backgrounds from the data itself (``off-source") and provides confidence in a signal should an excess be observed.

The main utility of this strategy will be in the search for 236~MeV neutrinos arising
from the leptonic decays of stopped kaons in the Sun.  Within 1 year of running, one would expect tens of $\sim \! 236\mev$ atmospheric neutrino background events at DUNE, making it desirable to reduce this background
in order to increase sensitivity~\cite{Rott:2015nma}.  On the other hand, with a similar exposure one would expect less than one background event arising from $\sim$~30~MeV atmospheric neutrinos;\footnote{Although the atmospheric neutrino background flux falls with energy, the neutrino-nucleus scattering cross section at the detector increases with energy.} so unless exposures are increased dramatically, directional information will be of little benefit in the search for 30 MeV neutrinos from stopped pions in the Sun.  In any case, the higher energy $236\mev$ neutrinos will impart a larger forward boost to a liberated  nucleon within the detector, improving one's ability to obtain direction information about the incoming neutrino.

We will use the \verb+NuWro+ software package~\cite{Golan:2012wx} to simulate charged-current interactions of $236\mev$ $\nu_e$ and $\nu_\mu$
with an argon target.  We find that a large fraction of events will, in addition to producing a charged lepton,
liberate a single proton from the nucleus, which otherwise remains intact.  This is a relatively clean signal
for a liquid argon time projection chamber (LArTPC) such as DUNE.  Moreover, since LArTPCs can reconstruct the track of the proton, we can use directional cuts on the proton's angular distribution to reduce the isotropic background neutrinos relative to the on-source signal neutrinos.  The \verb+NuWro+-generated event samples are used to determine these cut efficiencies.

It is important to keep in mind that the theoretical understanding of neutrino-nucleus interactions in the ${\cal O}(100)\mev$ energy range is far from complete.  As such, results such as cross sections and angular distributions obtained from any
computational tool may have only limited accuracy.  But we emphasize that these results really act as a proof-of-principle;
for an actual analysis, the angular cuts and associated efficiencies can be determined from calibrations which
can be performed at DUNE, as we will discuss.

This paper is structured as follows.  In section II, we will review the event rates at DUNE arising from dark
matter annihilation in the Sun, as well as from the atmospheric neutrino background.  In section III, we demonstrate
that one can significantly reduce the background at DUNE using cuts which preferentially select events where the
neutrinos arrive from the direction of the Sun.  We describe our results in section IV, and conclude in section V.

\section{The Neutrino Event Rate at DUNE}

We consider the scenario where dark matter collects in the core of the Sun after scattering against
solar nuclei, and then annihilates to a cascade of Standard Model particles whose subsequent decays
produce neutrinos.  We will be focused on the case of relatively low mass dark matter ($m_X \sim 10\gev$)
which exhibits spin-dependent scattering against protons, as this is the scenario in which the sensitivity
of neutrino detectors is most competitive with direct-detection experiments.  But we only consider the
mass range $m_X \gtrsim 4\gev$, as lighter dark matter will tend to evaporate from the Sun~\cite{WIMPevaporation}.
We assume here and throughout that the Sun is in equilibrium, so the total
dark matter annihilation rate ($\Gamma_A$) and the rate at which dark matter is captured by the Sun ($\Gamma_C$)
are related by $\Gamma_A =  \Gamma_C / 2$.  This is in fact a somewhat conservative assumption: for dark matter with a mass $m_X \sim 10 \gev$ and an
annihilation cross section $\langle \sigma_{ann.} v \rangle \sim 1\pb$, one would expect the Sun to be in equilibrium
if the dark matter-proton spin-dependent scattering cross section satisfies
$\sigmaSD^p > 3 \times 10^{-7}~\pb$~\cite{Kumar:2012uh}.

Let us assume that dark matter annihilates to light quark/anti-quark pairs.  These quarks will shower and hadronize to produce a number of long-lived mesons, which will subsequently undergo interactions with the dense nuclear medium of the Sun.  These interactions
will, in turn, result in further hadronic particle showers, yielding a very large number of secondary light mesons which will come to rest and decay.  In this way, the initial (high) energy released from dark matter annihilation is transformed into a large number of light mesons, whose decays at rest can produce low-energy neutrino signals that rise above the background. Almost any channel that yields high-energy neutrinos through decays will also be accompanied by this low-energy neutrino signal.  Consequently, should a signal be observed, the relationship between high- and low-energy signal contributions can potentially allow one to determine or constrain the mix of annihilation channels.

Of interest to us are $\pi^+$ and $K^+$, which decay via the process $\pi^+, K^+ \rightarrow \nu_\mu \mu^+$
with branching fractions of $\sim 100\%$ and $64\%$, respectively.  There will be no significant neutrino
signal arising from the decay of negatively charged mesons, such as $\pi^-$, as they will instead be
Coulomb-captured by nuclei \cite{Ponomarev:1973ya}.  There will also be no significant neutrino signal
from the neutral mesons, which predominately (and promptly) decay to photons.

The decay of a stopped $\pi^+$ or $K^+$ to the $\nu_\mu \mu^+$ final state will result in a monoenergetic
$\nu_\mu$ with energy of $29.8\mev$ or $235.5\mev$, respectively.  The subsequent decay of the $\mu^+$ will
produce continuum $\bar \nu_\mu$ and $\nu_e$ spectra as well, which have been subject to
previous studies~\cite{Rott:2012qb,Bernal:2012qh}.
For this analysis, we will focus on the monoenergetic $\nu_\mu$ from kaon decay.  After oscillations in the Sun and in vacuum, it can arrive at an Earth-based neutrino detector as a $\nu_e$ or $\nu_\mu$ with sufficient energy to produce a charged lepton\footnote{The process wherein a monoenergetic 235.5 MeV $\nu_\mu$ undergoes a charged-current interaction was previously studied in~\cite{Spitz:2014hwa}.} via \textit{quasi-elastic charged-current} (QECC) interactions of the form $\nu_\ell + n \rightarrow \ell^- + p$.
In fact, for energies below $1 \gev$, QECC interactions dominate over other possible interactions such as resonant production (e.g., $\nu_\ell + n \rightarrow \ell^{-}+\Delta^{+}$) or deep inelastic scattering.  Moreover, for $235.5 \mev$ incoming neutrinos the struck nucleon in QECC interactions is typically ejected from the target nucleus, and very little of the transferred energy remains with the remnant nucleus.  This is confirmed in Monte-Carlo simulations, but the fact that very little energy is transferred to the remnant nucleus is to be expected from the kinematics: For $2\! \rightarrow \! 2$ scatterings that do not liberate a nucleon, the amount of kinetic energy that can be transferred to nucleus is $\lesssim \mathcal{O} (3\mev)$.  One would then expect the recoil energy of the $\Ar{40}$ nucleus remnant to remain small even when a struck nucleon is liberated.  Moreover, Monte-Carlo simulations indicate that the small amount of energy that is transferred to the remnant nucleus will often excite the nucleus, and the energy released as the nucleus falls back to its ground state may be detected.  Thus at a detector with excellent energy resolution, one can reconstruct not only the energy of the charged lepton and ejected proton produced through these QECC interactions, but also the total energy of the incoming neutrino.

In determining the monoenergetic neutrino event rate at DUNE, we follow the notation and formalism
of~\cite{Rott:2015nma}.  The event rate may be expressed as
\bea
N_{S,B}^{e, \mu} &=&
T \int_{E_0 -\Delta E/2}^{E_0+\Delta E/2} dE \int d\Omega \,
\left[  \int dE' f(E,E') \, \frac{d^2 \Phi_{S,B}^{e, \mu} }{dE' d\Omega}
\times A_{\rm eff}^{e, \mu} (E') \times \eta_{S,B}^{e, \mu} (E') \right]  ,
\label{eq:numEvents_Formal}
\eea
where $E_0$ is the energy of the monoenergetic neutrino, and $\Delta E$ is the width of the energy
window over which one counts events.
$\Phi$ is neutrino flux at the detector, $T$ is the time exposure, and
$A_{\rm eff}$ is the effective area of the detector.  The efficiency with which events pass the selection cuts is denoted by $\eta$.
The superscripts $e$ and $\mu$ distinguish electron and muon neutrinos, while the subscripts $S$ and $B$ distinguish
signal and background.  Finally, $f(E,E')$ is a smearing function which accounts for the energy
resolution of the detector.

Although the effective area and efficiencies are energy-dependent, the energy dependence of the
atmospheric neutrino background flux at this energy range is only known to ${\cal O}(10\%)$, which is
comparable to the energy resolution.  For our purposes, then, we may approximate the effective area
and efficiencies to be constants, evaluated at the energy of the monoenergetic neutrino.  We then find
\bea
N_{S,B}^{e, \mu} &=&
f_{S,B} T  A_{\rm eff}^{e, \mu}  \eta_{S,B}^{e, \mu}
\left[ \int_{E_0-\Delta E/2}^{E_0+\Delta E/2} dE \int d\Omega \, \frac{d^2 \Phi_{S,B}^{e, \mu} }{dE d\Omega} \right]  ,
\label{eq:numEvents}
\eea
where
\bea
f_{S,B} &\equiv&  \left[ \int_{E_0-\Delta E/2}^{E_0+\Delta E/2} dE \int d\Omega \,
\left[  \int dE' f(E,E') \, \frac{d^2 \Phi_{S,B}^{e, \mu} }{dE' d\Omega} \right] \right]
\nonumber\\
&\,& \times
\left[ \int_{E_0-\Delta E/2}^{E_0+\Delta E/2} dE \int d\Omega
\frac{d^2 \Phi_{S,B}^{e, \mu} }{dE d\Omega}  \right]^{-1}
\eea
encapsulates the effects of the detector energy resolution.

The effective area can be expressed as
\bea
A_{\rm eff} &=& \sigma_{\nu \text{-Ar}} \times \frac{M_{\rm target}}{\kT} \times \frac{(6.022 \times 10^{23}) \times 10^9}{A_{\Ar{}}},
\nonumber\\
&=&  (5.1 \times 10^{-10}\m^2 )
\left( \frac{\sigma_{\nu \text{-Ar}}}{10^{-38}\cm^2 } \right) \left( \frac{M_{\rm target}}{34\kT} \right) ,
\label{eq:effArea}
\eea
where $A_{\Ar{}} \sim 39.95$ is the atomic mass of argon, $M_{\rm target}$ is the fiducial mass of the detector,
and $\sigma_{\nu \text{-Ar}}$ is the neutrino-argon scattering cross section.

Neutrino-nuclei cross sections are not well determined at these energies, either theoretically or experimentally.  It is common to calculate these cross sections and to model interaction events using Monte-Carlo techniques, and various software packages have been developed to this end.  In this work we use \verb+NuWro+, since this package utilizes the spectral function for $\Ar{40}$, which has been shown (\cite{Ankowski:2007wr}) to be more accurate at modeling interactions at these low energies than the Fermi-gas model typically used in neutrino event generators.  As with any major neutrino interaction simulation currently available, however, \verb+NuWro+ utilizes the \textit{impulse approximation}: in essence, the neutrino is assumed to interact with a single nucleon, which can subsequently interact with other nucleons within the nucleus.  At neutrino energies below $\approx 100\mev$ \cite{Ankowski:2007wr,Zmuda:2015twa}, however, the Compton wavelength associated with the initial interaction can become larger than the scale of the individual nucleons, and the impulse approximation may break down.  Since we are focusing on incoming neutrinos with energy $\approx 236 \mev$, uncertainties arising from the breakdown of the impulse approximation are assumed to be small; we will comment on the possibility of resolving these uncertainties through calibration in the Conclusions.

With \verb+NuWro+, we calculate the QECC interaction cross sections to be
\begin{align}
\begin{split}
\sigma_{\nu_e \text{-Ar}}^{QECC} &= 4.2 \times 10^{-38} \cm^2 ,
\\
\sigma_{\nu_\mu \text{-Ar}}^{QECC} &= 2.7 \times 10^{-38} \cm^2 .
\end{split}
\label{eq:xsecQECC}
\end{align}
We find that this is in general agreement (within $\sim 20\%$) 
with cross sections evaluated using the \verb+GENIE+ software package \cite{Andreopoulos:2009rq} (assuming a Fermi gas model for the nucleus),
as reported previously in \cite{Rott:2015nma}.  
We note here that the \textit{total} cross sections for $\nu_\ell + \Ar{40}$ are larger, since this interaction can proceed through other processes such as quasi-elastic neutral-current interactions or coherent scattering with the entire nucleus.  We find, however, that the fraction of non-QECC events that pass the event-selection cuts described below is nearly zero.  We thus choose to analyze only the QECC events; the efficiencies $\eta_{S,B}^{e,\mu}$ reported in this analysis represent fraction of QECC events that pass the selection and directional cuts.  Accordingly, in the definition of detector effective area, Eq.~\eqref{eq:effArea}, we use these cross sections for purely QECC scattering.

\subsection{The Rate of Signal Events}

The neutrino flux arising from dark matter annihilation in the Sun can be expressed as
\bea
\frac{d^2 \Phi_S^{e, \mu}}{dE d\Omega} &=& \frac{(\Gamma_C / 2)F^{e,\mu} }{4\pi r_{\oplus}^2 } \left( 0.64 \times \frac{2m_X}{m_K} r_{K}(m_X) \right)
\delta (E-E_0) \delta(\Omega),
\eea
where $\Gamma_C$ is the rate at which dark matter is captured by the Sun, $F^{e,\mu}$ is the fraction of the $\nu_\mu$ produced by
stopped $K^+$ decay which arrive at the detector as either $\nu_e$ or $\nu_\mu$, and $r_{K}(m_X)$ is the fraction of the center-of-mass energy
in the annihilation process which goes into stopped kaons as a result of hadronization and subsequent nuclear processes in the Sun.
The factor of $0.64$ corresponds to the branching fraction for the process $K^+ \rightarrow \nu_\mu \mu^+$.
The Earth-Sun distance is $r_\oplus \sim 1.5 \times 10^{11}\m$, and the $\delta$-functions enforce the conditions that the flux be
of monoenergetic neutrinos emanating from the core of the Sun.  We then find
\bea
f_S &=& \int_{E_0-\Delta E/2}^{E_0+\Delta E/2} dE \, f(E,E_0) .
\eea
We will chose the energy width over which we sum events to be given by $\Delta E = \epsilon E_0$,
where $\epsilon$ is the fractional full-width energy resolution.  In this case,
$f_S \sim 0.68$.
We will discuss the possibility of other choices for $\Delta E$ in the Conclusions.

For an overview of the calculation of the dark matter capture rate, we refer the reader to~\cite{Gould:1987ir,Gould:1991hx}.
We point out that dark matter capture rates in the Sun show little dependence on the underlying assumptions
on the dark matter velocity distribution and other astrophysical uncertainties~\cite{Choi:2013eda,Danninger:2014xza}.
Assuming the standard local dark matter density and that dark matter-nucleon scattering is spin-dependent,
the capture rate can be expressed as $\Gamma_C = C_0^{SD} (m_X) \times \sigmaSD^p $, where $\sigmaSD^p$ is the
spin-dependent dark matter-proton scattering
cross section and the $C_0^{SD} (m_X)$ are coefficients which can be found, for example, in~\cite{Gao:2011bq,Kumar:2012uh}.  The $r_{K}(m_X)$ were
determined in~\cite{Rott:2015nma}, and the $F^{e,\mu}$ can be found in~\cite{Lehnert:2007fv}.  In particular, assuming a normal hierarchy
(which we do henceforth) and
$E_0 = 236\mev$, we find that $F^e = 0.46$ and $F^\mu =0.27$.  With these pieces, one can determine the rate of signal events as a function of $m_X$ and
$\sigmaSD^p$ for any choice of exposure and $\eta_S$.

\subsection{The Rate of Background Events}

The dominant source of background events will be atmospheric neutrinos.  The atmospheric neutrino background fluxes at
$236\mev$ are given by~\cite{Battistoni:2005pd}
\begin{align}
\begin{split}
\frac{d^2 \Phi_B^e}{dE d\Omega} &\sim 1.2 ~\m^{-2} \s^{-1} \sr^{-1} \mev^{-1} ,
\\
\frac{d^2 \Phi_B^\mu}{dE d\Omega} &\sim 2.3 ~\m^{-2} \s^{-1} \sr^{-1} \mev^{-1} .
\end{split}
\end{align}
If we approximate the background atmospheric neutrino flux to be constant over this energy range, then we
find
\bea
f_B = \int dE' \, f(E,E') =1.
\eea

An additional source of background events arises from atmospheric \textit{anti}-neutrinos, which have fluxes of roughly the same magnitude.  These will typically scatter through QECC interactions of the form $\nu_\ell p \rightarrow \ell^{+} n$, and for a small number of events the outgoing neutron will cause a proton to be ejected through intra-nuclear interactions.  Since the LArTPCs cannot typically distinguish between positive and negative charged leptons, some of these events will pass our selection cuts.  Through \verb+NuWro+ simulation we determine that these events are negligible to this analysis, as they contribute less than $1\%$ of total events passing our cuts.

The number of background events, for any given exposure, is thus
\begin{align}
\begin{split}
N_B^e &= 240 \left(\frac{\rm exposure}{34\kT \yr}\right) \times \epsilon \times \eta_B^e ,
\\
N_B^\mu &= 300 \left(\frac{\rm exposure}{34\kT \yr}\right) \times \epsilon \times \eta_B^\mu ,
\end{split}
\end{align}
where again we have set $\Delta E = \epsilon E_0$.
We note here that the atmospheric neutrino flux depends on the location of the detector, and may vary from the ones used here by up to a factor of two~\cite{Battistoni:2005pd}.  The uncertainty in this analysis can be substantially reduced by a more precise calculation of the atmospheric neutrino flux at the specific location of DUNE, but such a calculation is beyond the scope of this work.

\section{Event Selection and Directional Cuts}

The overall cut efficiencies $\eta_{S,B}$ introduced in Eq.~\eqref{eq:numEvents} can be decomposed into a product of two factors, $\eta_{S,B} = \eta_\text{sel}  \cdot \eta_{\text{dir}(S,B)}$.  The factor $\eta_\text{sel}$ represents the fraction of neutrino scattering events which pass the event selection and detector threshold cuts.  This factor will be identical for both signal and background neutrinos.  The second factor $\eta_{\text{dir}(S,B)}$ represents the fraction of events which pass the directionality cuts; since the atmospheric neutrino background is largely isotropic while the signal neutrinos originate from the Sun, we will in general have $\eta_{\text{dir} (S)} \geq \eta_{\text{dir} (B)}$.

In order to determine the overall cut efficiencies $\eta_{S,B}$, we analyze $10^5$ QECC $\nu_\ell + \Ar{40}$ events generated in \verb+NuWro+, for both electron and muon incoming neutrinos of energy $E_\nu = 235.5 \mev$.
In order to use the $\Ar{40}$ spectral function, we set \verb+nucleus_target=2+ and \verb+sf_method=1+ in the \verb+NuWro+ parameter-initialization file.  All other parameters are set to default values.  In addition, we have confirmed that our results are insensitive to small variations of the simulation settings, such as quasielastic form-factor coefficients or final state interaction parameters.

\begin{figure}[h!]
  \centering
  \includegraphics[width= 0.6 \textwidth]{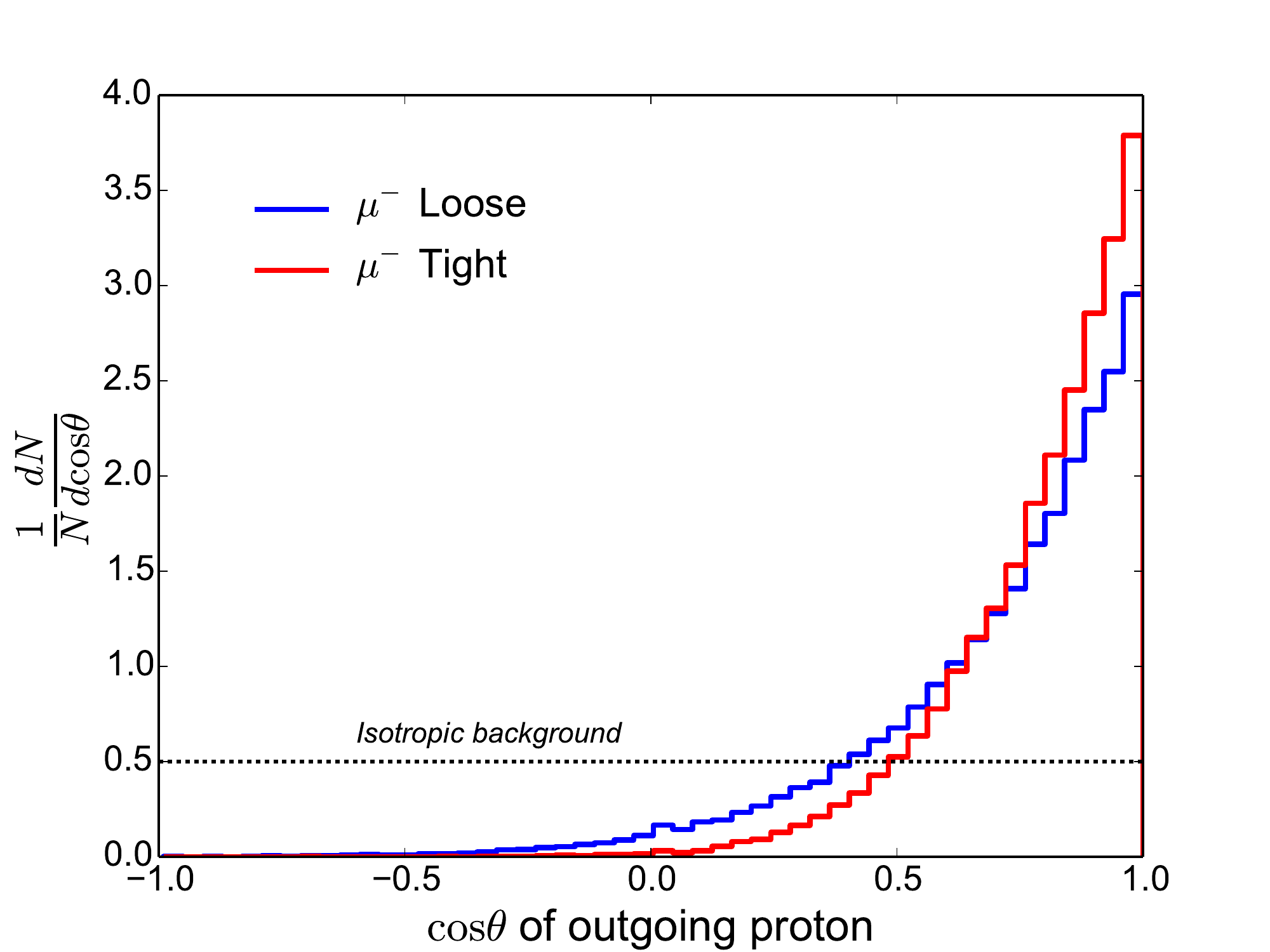}
  \cprotect\caption{\label{fig:CosTheta} Number density of $\nu_{\mu} + \Ar{40}$ events which pass selection cuts as a function of angle $\theta$ of the outgoing proton, showing that protons are typically ejected in the forward direction.  The angle is measured from the direction vector of the incoming neutrino.  The small number of multi-nucleon knockout events which pass selection cuts are included in these distributions.  We have overlaid the number densities for two different choices of detector energy thresholds, which are discussed in the text.  These distributions are generated in \verb+NuWro+, and similar distributions for the $\nu_e$ channel are essentially identical.}
\end{figure}

We focus on charged-current interactions, because they will produce a charged lepton which can be easily measured
by the detector.
As mentioned previously, $235.5 \mev$ neutrinos will dominantly scatter through QECC interactions with nucleons
within the target nuclei, $\nu_{\ell} + n \rightarrow \ell^- + p^+$.
LArTPC detectors such as DUNE are well-suited to detect the energy of both the charged lepton and the ejected proton, as well as the total energy of the incoming neutrino.  Moreover, since the proton will typically be ejected from the nucleus in the forward direction (see Fig.~\ref{fig:CosTheta}), LArTPCs with good angular resolution would have directional sensitivity to the incoming neutrino.  For these reasons, we require events to contain a single proton and a single charged lepton in the final state.

In conjunction with this event selection requirement, we must also take into account the sensitivity of the detector.  In order
for either lepton or proton track to be identified, the kinetic energy $E_\text{kin}$ must be greater than a minimum
threshold.  For the charged lepton, we will take this threshold to be $30\mev$.
For the proton, a conservative estimate for the identification threshold is $50\mev$~\cite{Acciarri:2015uup};
we will refer to this requirement as the ``tight'' threshold in what follows.  We will also consider a more optimistic ``loose'' threshold of $20\mev$.

Our selection cuts thus require that events have exactly one \textit{visible} (as in, above threshold $E_\text{kin}$) proton and one visible charged lepton.  The fraction of total QECC events that pass these cuts is equal to $\eta_\text{sel}$.  These efficiencies are collected in Table.~\ref{tab:SelectionCutEfficiency}.    We mention here that we have imposed cuts only at the level of event generation, and have made no attempt to realistically simulate the detector response.

In addition single proton ejection events of interest to us here, there will be some number of QECC events where either zero nucleons or more than one nucleon are ejected from the remnant nucleus.  These events are mainly the result of intranuclear processes, where the primary struck nucleon interacts with and transfers energy to other nucleons within the nucleus.  A small number of multi-nucleon knockout events --- where more than one proton or where one or more neutrons are ejected --- will pass the selection cuts.  This will be the case, for instance, when a scattering event ejects two protons, with only one of these protons above the kinetic energy threshold; Events where an above-threshold proton and a neutron are ejected will also pass our selection requirement, since typically the neutron will escape detection.  In these cases, there will be less of a correlation between the tagged proton direction and incoming neutrino direction, since there will be missing momentum associated with the untagged nucleon(s).

The final state interactions which give rise to zero- or multi-nucleon knockout QECC events are modeled in \verb+NuWro+ using the intra-nuclear cascade (INC) approach (see \cite{Golan:2012wx} for details concerning accuracy and uncertainty in this modeling).  All results presented here, including the efficiencies $\eta_\text{sel}$ and $\eta_\text{dir}$, are derived from \verb+NuWro+ event samples which include this FSI modeling.  Although the \verb+NuWro+ INC approach is expected to be reliable, we mention here that simply turning off all FSI modeling has a less than 5\%  affect on these efficiencies.  Moreover, for events passing selection cuts, the normalized proton angular distributions of events generated without FSI are essentially indistinguishable from the distributions seen in Fig.~\ref{fig:CosTheta}.  Thus the main results demonstrated here --- that directionality can be used in these cases to reduce backgrounds --- are largely insensitive to the details of how intranuclear modeling is performed.

\begin{table}[h]
\centering
\begin{tabular}{|c|c|c|}
  \hline
  cut & proton threshold  & selection efficiency ($\eta_\text{sel}$) \\
  \hline
  tight: electron & $E_\text{kin} > 50\mev$ & $0.43$ \\
  tight: muon & $E_\text{kin} > 50\mev$ & $0.28$ \\
  \hline
  loose: electron & $E_\text{kin} > 20\mev$ & $0.83$ \\
  loose: muon & $E_\text{kin} > 20\mev$ & $0.75$ \\
  \hline
\end{tabular}
\caption{The fraction of QECC events which result in the production of a single charged lepton with
$E_\text{kin} > 30\mev$ and a single ejected proton satisfying the listed cuts. Note that these efficiencies include multi-nucleon knockout events which pass the selection criteria. \label{tab:SelectionCutEfficiency}}
\end{table}

We can now impose a further directional cut, requiring the proton to be ejected within a cone of angle $\theta$ centered on the direction from the Sun.
The ratio of the selected events passing this directionality cut determines the efficiency $\eta_\text{dir}$.
Since the protons are typically ejected in the forward direction, this cut can significantly reduce the number of background events relative to signal events.  We will assume that, for the energy regime of interest, the angular resolution of LArTPCs to the proton track is $5^\circ$~\cite{Acciarri:2015uup}.  Note that there is little to be gained from a similar cut on the lepton direction, as the leptons are produced largely isotropically at these energies.

\begin{figure}[h!]
  \centering
  \includegraphics[width= 0.6 \textwidth]{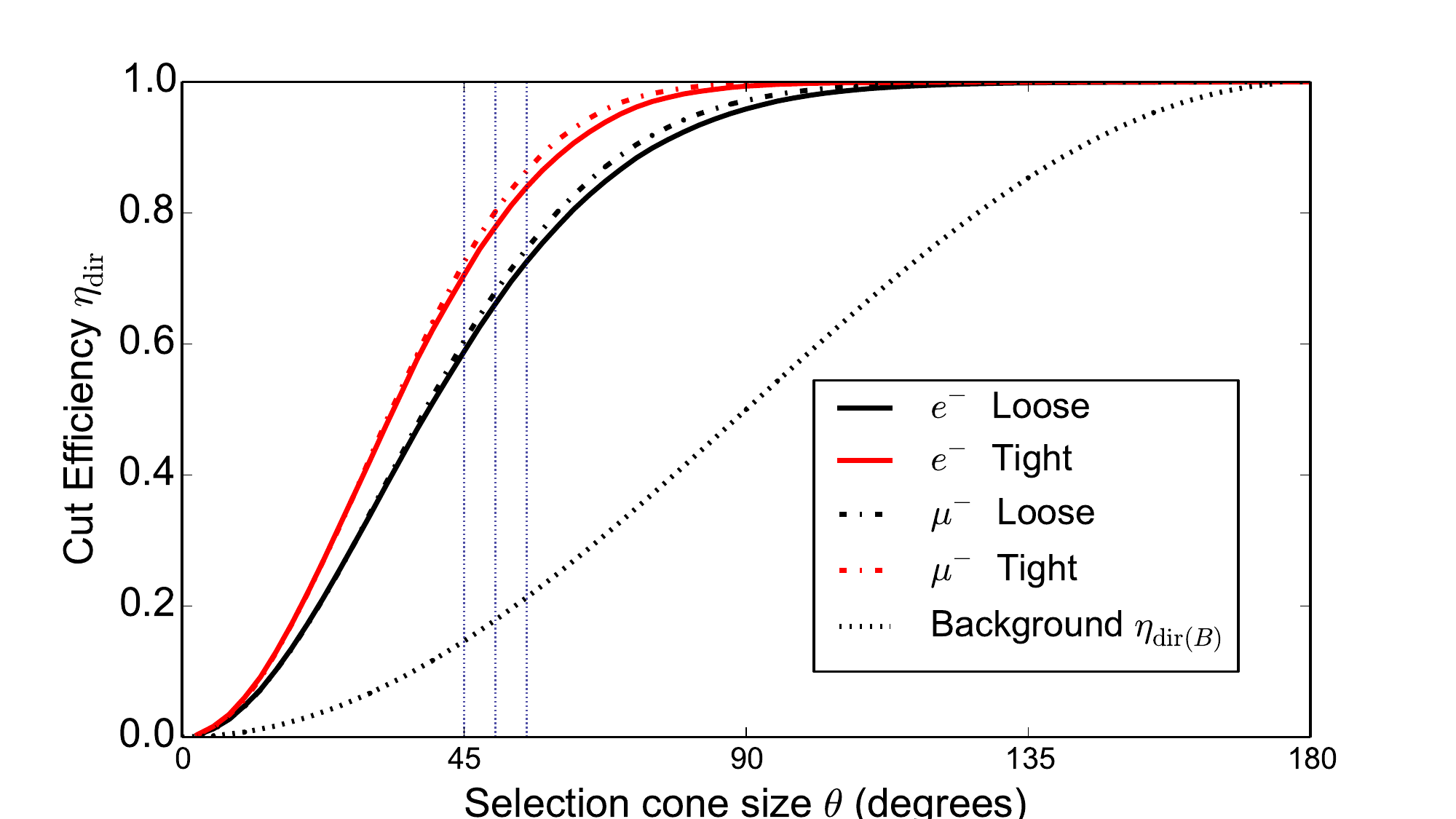}
  \caption{\label{fig:DirCutEff} Directionality cut efficiencies $\eta_{\text{dir}(S)}^{e,\mu}$ for tight and loose signal events.  Also shown is the cut efficiency on background $\eta_{\text{dir}(B)}$, which simply shows isotropically distributed protons within cone size $\theta$.  The vertical lines correspond to the angles which maximize the signal statistical significance.}
\end{figure}

Since the atmospheric neutrino background is largely isotropic, the fraction of background events that will satisfy this cut is roughly
\begin{equation}
\eta_{\text{dir}(B)} = \tfrac{1}{2} \left[ 1-\cos (\theta/2) \right].
\end{equation}
For the signal, the directionality cut efficiency $\eta_{\text{dir}(S)}$ is calculated from the $\eta_\text{sel} \times 10^5$ \verb+NuWro+-generated events which have already passed the selection cuts.  These events are generated assuming an incoming neutrino beam traveling in the $z$-direction; we simply count the number of (selected) events which produce a proton within a cone centered on the positive $z$-axis.  The directionality cut efficiency calculated in this manner is shown in Fig.~\ref{fig:DirCutEff} for both incoming $\nu_e$ and $\nu_\mu$, and for events selected using both ``loose'' and ``tight'' selection cuts.
The total cut efficiency $\eta_{S,B}^{e,\mu}$ is now simply given by multiplying the selection efficiencies $\eta_\text{sel}$
in Table~\ref{tab:SelectionCutEfficiency} by the directionality cut efficiencies $\eta_{\text{dir}(S,B)}$ in Fig.~\ref{fig:DirCutEff}.

We will choose the angular cut on the direction of the proton in order to maximize the improvement in statistical
significance.  In the limit in which Gaussian statistics are appropriate, this amounts to maximizing the quantity
$\eta_S / \sqrt{\eta_B}$.  We list the relevant cuts and efficiencies in Table~\ref{tab:Cuts}.  Note, for all of the
directional cuts, the half-angle is much larger than the angular resolution.  We have not accounted for the smearing of the
reconstructed proton direction due to the angular resolution, because this is in any case a very small effect compared
to the other uncertainties in the problem.

For the case of
loose cuts on the proton kinetic energy ($E_{kin} > 20\mev$), the quantity $S / \sqrt{B}$ can be increased by a
factor of $\sim \! 40\%$ by the use of directionality cuts.  But for the case of tight cuts on the proton energy
($E_{kin} > 50\mev$), $S / \sqrt{B}$ is essentially unchanged.
Although the signal significance is only marginally improved, the signal-to-background ratio always increases dramatically, by a
factor of $\sim \! 4-5$ for tight cuts, and $\sim \! 3.5$ for loose cuts.  Moreover, in all cases, $\sim \! 70\text{--}80\%$ of signal
events are expected to pass the directional cuts, compared to $\sim \! 15\text{--}20\%$ of background events.  The striking
difference between ``on-source" and ``off-source" event rates will be useful in distinguishing signal from background.
\begin{table}[h]
\centering
\begin{tabular}{|c|c|c|c|c|c|}
  \hline
  cut & half-angle & $\eta_S$ & $\eta_B$ & $S/B$ & sensitivity \\
   & & & & enhancement & enhancement \\
  \hline
  tight: electron & $45^\circ$ & $\eta_S^e = 0.30$ & $\eta_B^e = 0.06$ & 5.0 & 1.2 \\
  tight: muon & $50^\circ$ & $\eta_S^\mu = 0.23$ & $\eta_B^\mu = 0.05$ & 4.6 & 1.0 \\
  \hline
  loose: electron & $55^\circ$ & $\eta_S^e = 0.60$ & $\eta_B^e = 0.18$ & 3.3 & 1.4 \\
  loose: muon & $55^\circ$ & $\eta_S^\mu = 0.56$ & $\eta_B^\mu = 0.16$ & 3.5 & 1.4 \\
  \hline
\end{tabular}
\caption{ The cone half-angle, in the direction from the Sun, within which the ejected
proton track must lie for each of the listed cuts.  These cone-angles maximize the increase in statistical significance gained by using this directional search strategy.  Also given is the total efficiency
of each set of cuts for signal ($\eta_S$) and background ($\eta_B$) events, as well
as the enhancement to the signal-to-background ratio obtained by applying each
set of cuts.
The last column gives the factor by which sensitivity is enhanced, for a fixed exposure, by
the application of the given directional cuts.
\label{tab:Cuts}}
\end{table}

\section{Results}

\begin{figure}[ht]
\centering
\includegraphics[scale=0.60]{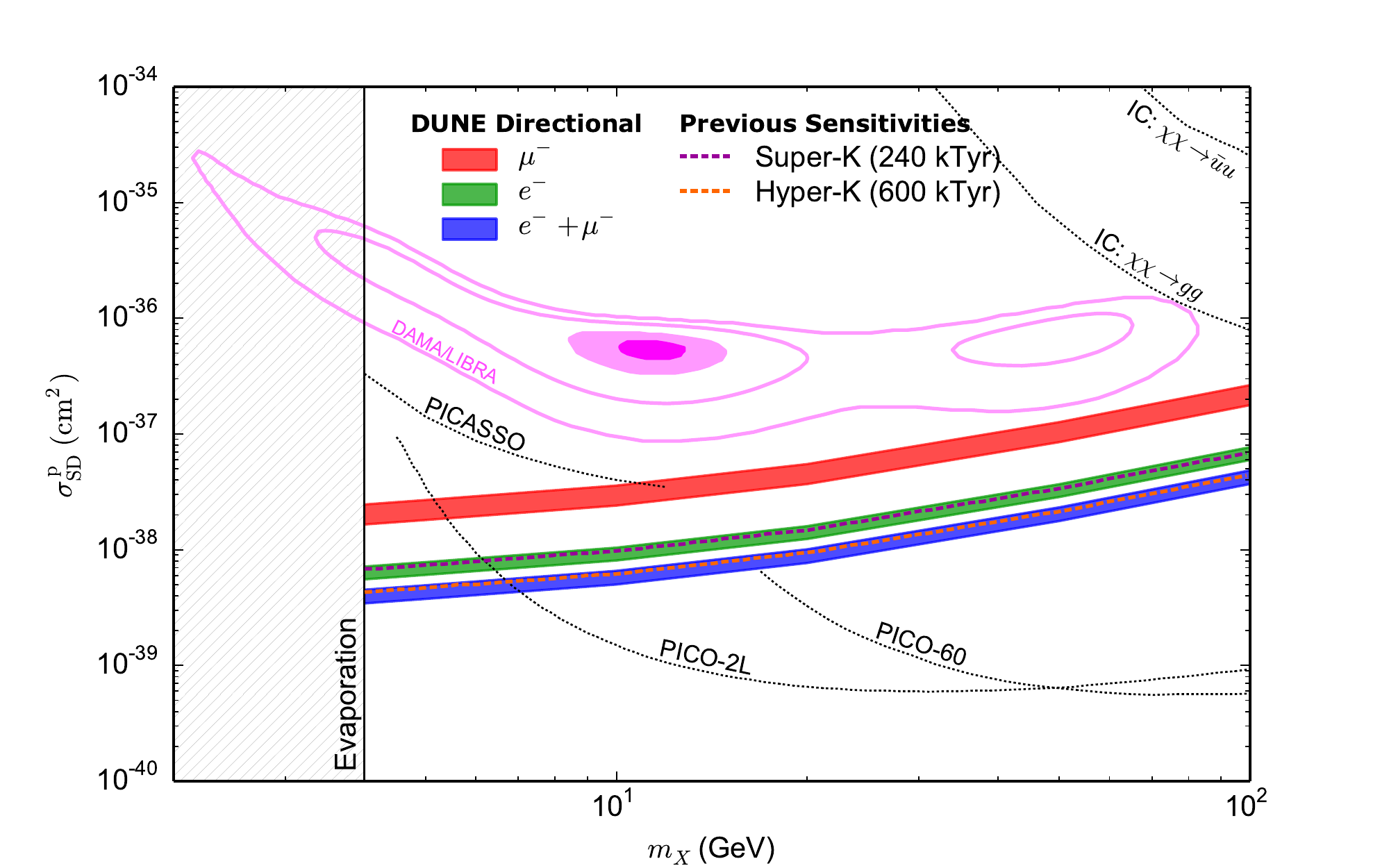}
\cprotect\caption{Experimental sensitivity of DUNE at 90\%CL (with an exposure of 340 kT yr) using the directional search described in this work.  The bands span the sensitivity probed using the tight proton energy threshold requirement (upper edges) and the more optimistic loose threshold energy requirement (lower edges).  The blue band represents the combined sensitivities from the electron-only and muon-only analysis.
Also shown are previous sensitivities from DUNE~\cite{Acciarri:2015uup}, Super-K~\cite{Fukuda:2002uc}, and Hyper-K~\cite{Abe:2015zbg,Abe:2011ts} when directional information is not used, as found in~\cite{Rott:2015nma}.  We also show the exclusion regions based on IceCube data~\cite{Aartsen:2016exj,Aartsen:2012kia} computed with \verb+nulike+ ~\cite{nulike,Aartsen:2016exj,Scott:2012mq}, on PICO-60~\cite{Amole:2015pla} and on PICO-2L~\cite{Amole:2016pye},
and the region favored by DAMA/LIBRA (at $90\%/3\sigma/5\sigma/7\sigma$ CL)~\cite{Savage:2008er}.
\label{fig:ExclusionContour}}
\end{figure}

We will consider, as an example, a sensitivity estimate for DUNE running with an exposure of 340 kT yr
and a 10\% energy resolution ($\epsilon=0.1$).
As a benchmark, we consider the scenario where dark matter collects in the core of the Sun as a result of
spin-dependent scattering with nuclei in the Sun, and then annihilates to first-generation quarks (the average number of
$K^+$ produced per annihilation is the same for the $\bar u u$ and $\bar d d$ channels~\cite{Rott:2015nma}).
We consider muon and electron channels separately to compare their performance, and then combine them to provide the optimal sensitivity.
For simplicity we assume that the number of observed electron or muon events ($N_{obs}$) is equal to the number
of expected background events, rounded to the nearest integer. A signal is excluded at $90\%$CL if the Poisson-distributed total number of events consisting of the sum of background and signal exceeds the assumed observed number of events in at least $90\%$ of the cases.
For each set of cuts, the number of expected background events, assumed observed events, and the number of expected
signal events needed for $90\%$CL exclusion is presented in Table~\ref{tab:NumberOfEvents}.
We also determine the sensitivity of a joint electron/muon analysis; in this case, a cross section is excluded at 90\%CL if the probability of observing $N_{obs}^e$ or fewer electron events given the expected number of electron events ($N_{S}^e +N_{B}^e$), multiplied by the probability of observing $N_{obs}^\mu$ or fewer muon events given the expected number of muon events, is at most 10\%.
The corresponding cross section exclusion contours, as a function of $m_X$, are given in Figure~\ref{fig:ExclusionContour}.
These contours are represented by bands, with the upper and lower edges of the bands corresponding to the tight and loose threshold energy requirements, respectively.

\begin{table}[h]
\centering
\begin{tabular}{|c|c|c|c|}
  \hline
  cuts & expected $N_B$ & assumed $N_{obs}$ & expected $N_S$ for exclusion \\
  \hline
  tight: electron & 14.8 & 15 & 6.5 \\
  tight: muon & 14.9 & 15 & 6.4 \\
\hline
  loose: electron & 41.6 & 42 & 10.0 \\
  loose: muon & 47.5 & 48 & 10.7 \\
  \hline
\end{tabular}
\caption{ The number of expected background events ($N_B$) which would pass each set of cuts at DUNE,
assuming $\epsilon = 0.1$ and a 340 kT yr exposure.  Also given are the number of events assumed to
observed ($N_{obs}$) for the purpose of this analysis, and the number of expected signal events passing
the cuts which would be required in order for the model to be excluded at $90\%$CL.
\label{tab:NumberOfEvents}}
\end{table}

The constraints from the electron channel are tighter than those from the muon channel.  This is due to a combination of three effects:
the greater effective area of the detector for the electron channel, the larger electron neutrino flux from the Sun resulting from oscillation
effects, and the smaller
atmospheric electron neutrino background.  Moreover, the loose cuts provide for slightly better sensitivity than the tight cuts, because
of the greater efficiency of the selection cuts.  Although these cuts provide a marginal improvement in
statistical sensitivity, the real advantage lies in the improved signal-to-background ratio; as we see from Table~\ref{tab:NumberOfEvents},
the signal-to-background ratio ($N_S/N_B$) can be as large as $43\%$ in the case of tight cuts (electron or muon), an improvement by a factor of
$\sim 4-5$ over the case without these cuts.
This is especially important because the signal is heavily peaked on-source, allowing one to
more easily distinguish signal from an unexpected background.

We have assumed an energy resolution of $10\%$ and that energy and angular resolutions do not depend on
the direction of the incoming neutrino flux.  To rescale these limits for other choices of the energy resolution, we note that, in
the limit where background is significant, the sensitivity scales as $\epsilon^{-1/2}$.  To determine the exact
sensitivity for DUNE a full detector simulation is required, which is beyond the scope of this work. We encourage the
DUNE collaboration to carry out such a study.

\section{Conclusions}

We have considered the possibility of using DUNE to perform a directional search for the monoenergetic $236\mev$ neutrinos
which can result from the copious decays of stopped $K^+$ produced by dark matter annihilation in the core of the Sun.  When
such low-energy neutrinos interact with the nuclei in the detector through a charged-current interaction,
there is often an ejected proton with a forward-peaked angular
distribution.  Although the gain in statistical significance when using this directional information is only marginal,
the increase in the signal-to-background ratio is substantial.  The greatest utility of this technique thus lies in reducing
systematic uncertainties in studies of low-energy neutrinos arising from dark matter annihilation in the Sun.  Although water Cherenkov
neutrino detectors, such as Super-Kamiokande and Hyper-Kamiokande~\cite{Abe:2015zbg,Abe:2011ts}, may have greater sensitivity to ${\cal O}(100)\mev$ neutrinos
due to their extremely large exposures, they are not capable of performing this type of a directional search.  Large LArTPC detectors
such as DUNE thus have a unique ability to perform directional neutrino searches at this energy range.

Our assumptions about the thresholds for proton identification at DUNE are based on preliminary estimates.  It is clear that
lower thresholds can potentially increase the statistical power of this search strategy.
As we previously noted, there is a great deal of theoretical uncertainty regarding the neutrino-nucleus scattering
cross section in the ${\cal O}(100)\mev$ energy range.  As such, one should best treat these results as a proof-of-principle
regarding the feasibility of obtaining direction information for sub-GeV neutrinos at DUNE.  For an actual dark matter
search, it would be desirable to calibrate the detector with a dedicated stopped kaon experiment.  Any stopped pion experiment
is also a stopped kaon experiment~\cite{Spitz:2012gp}, and experiments of this type, such as DAE$\delta$AELUS~\cite{Conrad:2010eu} are planned for DUNE.

In this vein, we note that a beam dump experiment in the vicinity of DUNE could be used to precisely determine the exact signal efficiency for a given cone opening angle and energy window. With the exception of some dependence on the neutrino flavor ratio the signal efficiency can hence be precisely measured. Backgrounds can be determined from off-source regions and and therefore can be precisely determined. In combination, this would allow one to optimize the choice of energy and angular windows and perform a very robust analysis.

Finally, we point out that although we have studied directional searches at DUNE in the context of the $236\mev$ monoenergetic
neutrino which can be produced by stopped $K^+$ decay in the Sun, this technique can be applied more broadly to any directional
signal of ${\cal O}(100)\mev$ neutrinos.  This technique thus has much broader applicability.

\vskip .2in
\acknowledgments

We are grateful to Danny Marfatia, Jelena Maricic, Jan Sobczyk, and Elizabeth Worcester for useful discussions.
We would like to thank Pat Scott and Matthias Danninger for computing light quark bounds with \verb+nulike+. The work of J.~Kumar is
supported in part by NSF CAREER Grant No.~PHY-1250573.  C.~Rott  acknowledges support from the Korea Neutrino Research Center which is established by the National Research Foundation of Korea (NRF) grant funded by the Korea government (MSIP) (No. 2009-0083526) and Basic Science Research
Program NRF-2016R1D1A1B03931688.  S.~In is supported by Global PH.D Fellowship Program through the National Research Foundation of Korea (NRF) funded by the Ministry of Education (NRF-2015H1A2A1032363). D.~Yaylali is supported in part by DOE grant DE-FG02-13ER-41976.
J.~Kumar would like to thank CETUP* (Center for Theoretical Underground Physics and Related Areas), for hospitality and partial support.






\begin{thebibliography}{99}
\label{bibs}



\bibitem{Silk:1985ax}
  J.~Silk, K.~A.~Olive and M.~Srednicki,
  Phys.\ Rev.\ Lett.\  {\bf 55}, 257 (1985).

  \bibitem{Press:1985ug}
  W.~H.~Press and D.~N.~Spergel,
  Astrophys.\ J.\  {\bf 296}, 679 (1985).

  \bibitem{Krauss:1985ks}
  L.~M.~Krauss, K.~Freese, W.~Press and D.~Spergel,
  Astrophys.\ J.\  {\bf 299}, 1001 (1985).

\bibitem{Aartsen:2012kia}
  M.~G.~Aartsen {\it et al.} [IceCube Collaboration],
  Phys.\ Rev.\ Lett.\  {\bf 110}, no. 13, 131302 (2013)
  [arXiv:1212.4097 [astro-ph.HE]].

\bibitem{Choi:2015ara}
  K.~Choi {\it et al.} [Super-Kamiokande Collaboration],
  Phys.\ Rev.\ Lett.\  {\bf 114}, no. 14, 141301 (2015)
  [arXiv:1503.04858 [hep-ex]].

\bibitem{Adrian-Martinez:2013ayv}
  S.~Adrian-Martinez {\it et al.} [ANTARES Collaboration],
  JCAP {\bf 1311}, 032 (2013)
  [arXiv:1302.6516 [astro-ph.HE]].



\bibitem{Rott:2012qb}
  C.~Rott, J.~Siegal-Gaskins and J.~F.~Beacom,
  Phys.\ Rev.\ D {\bf 88}, 055005 (2013)
  [arXiv:1208.0827 [astro-ph.HE]].

\bibitem{Bernal:2012qh}
  N.~Bernal, J.~Martín-Albo and S.~Palomares-Ruiz,
  JCAP {\bf 1308}, 011 (2013)
  [arXiv:1208.0834 [hep-ph]].

\bibitem{Kumar:2015nja}
  J.~Kumar and P.~Sandick,
  JCAP {\bf 1506}, no. 06, 035 (2015)
  [arXiv:1502.02091 [hep-ph]].

\bibitem{Rott:2015nma}
  C.~Rott, S.~In, J.~Kumar and D.~Yaylali,
  JCAP {\bf 1511}, no. 11, 039 (2015)
  [arXiv:1510.00170 [hep-ph]].

\bibitem{Acciarri:2015uup}
  R.~Acciarri {\it et al.} [DUNE Collaboration],
  arXiv:1512.06148 [physics.ins-det].

\bibitem{Kim:2014rfa}
  S.~B.~Kim,
  Nucl.\ Part.\ Phys.\ Proc.\  {\bf 265-266}, 93 (2015)
  doi:10.1016/j.nuclphysbps.2015.06.024
  [arXiv:1412.2199 [hep-ex]].

\bibitem{Djurcic:2015vqa}
  Z.~Djurcic {\it et al.} [JUNO Collaboration],
  arXiv:1508.07166 [physics.ins-det].

\bibitem{Alonso:2014fwf}
  J.~R.~Alonso {\it et al.},
  arXiv:1409.5864 [physics.ins-det].

\bibitem{Abe:2015zbg}
  K.~Abe {\it et al.} [Hyper-Kamiokande Proto- Collaboration],
  PTEP {\bf 2015}, no. 5, 053C02 (2015)
  [arXiv:1502.05199 [hep-ex]].

\bibitem{Abe:2011ts}
  K.~Abe {\it et al.},
  arXiv:1109.3262 [hep-ex].



\bibitem{Golan:2012wx} 
  T.~Golan, C.~Juszczak and J.~T.~Sobczyk,
  Phys.\ Rev.\ C {\bf 86}, 015505 (2012)
  [arXiv:1202.4197 [nucl-th]].

\bibitem{WIMPevaporation}
  K.~Griest and D.~Seckel,
  Nucl.\ Phys.\ B {\bf 283}, 681 (1987)
  [Erratum-ibid.\ B {\bf 296}, 1034 (1988)];
  A.~Gould,
  Astrophys.\ J.\  {\bf 321}, 560 (1987).

\bibitem{Kumar:2012uh}
  J.~Kumar, J.~G.~Learned, S.~Smith and K.~Richardson,
  Phys.\ Rev.\ D {\bf 86}, 073002 (2012)
  [arXiv:1204.5120 [hep-ph]].

\bibitem{Ponomarev:1973ya}
  L.~I.~Ponomarev,
  Ann.\ Rev.\ Nucl.\ Part.\ Sci.\  {\bf 23}, 395 (1973).

\bibitem{Spitz:2014hwa}
  J.~Spitz,
  Phys.\ Rev.\ D {\bf 89}, no. 7, 073007 (2014)
  [arXiv:1402.2284 [physics.ins-det]].

\bibitem{Ankowski:2007wr}
  A.~M.~Ankowski and J.~T.~Sobczyk,
  AIP Conf.\ Proc.\  {\bf 967}, 106 (2007)
  [arXiv:0709.2139 [nucl-th]].



\bibitem{Zmuda:2015twa}
  J.~\.{Z}muda, K.~M.~Graczyk, C.~Juszczak and J.~T.~Sobczyk,
  Acta Phys.\ Polon.\ B {\bf 46}, no. 11, 2329 (2015)
  [arXiv:1510.03268 [hep-ph]].

\bibitem{Andreopoulos:2009rq}
  C.~Andreopoulos {\it et al.},
  Nucl.\ Instrum.\ Meth.\ A {\bf 614}, 87 (2010)
  [arXiv:0905.2517 [hep-ph]].

\bibitem{Gould:1987ir}
  A.~Gould,
  Astrophys.\ J.\  {\bf 321}, 571 (1987).

\bibitem{Gould:1991hx}
  A.~Gould,
  Astrophys.\ J.\  {\bf 388}, 338 (1992).

\bibitem{Choi:2013eda}
  K.~Choi, C.~Rott and Y.~Itow,
  JCAP {\bf 1405}, 049 (2014)
  [arXiv:1312.0273 [astro-ph.HE]].

\bibitem{Danninger:2014xza}
  M.~Danninger and C.~Rott,
  Phys.\ Dark Univ.\  {\bf 5-6}, 35 (2014)
  [arXiv:1509.08230 [astro-ph.HE]].



\bibitem{Gao:2011bq}
  Y.~Gao, J.~Kumar and D.~Marfatia,
  Phys.\ Lett.\ B {\bf 704}, 534 (2011)
  [arXiv:1108.0518 [hep-ph]].

\bibitem{Lehnert:2007fv}
  R.~Lehnert and T.~J.~Weiler,
  Phys.\ Rev.\ D {\bf 77}, 125004 (2008)
  [arXiv:0708.1035 [hep-ph]].


\bibitem{Battistoni:2005pd}
  G.~Battistoni, A.~Ferrari, T.~Montaruli and P.~R.~Sala,
  Astropart.\ Phys.\  {\bf 23}, 526 (2005).


\bibitem{Fukuda:2002uc}
  Y.~Fukuda {\it et al.} [Super-Kamiokande Collaboration],
  Nucl.\ Instrum.\ Meth.\ A {\bf 501}, 418 (2003).


\bibitem{Aartsen:2016exj}
  M.~G.~Aartsen {\it et al.} [IceCube Collaboration],
  JCAP {\bf 1604}, no. 04, 022 (2016)
  [arXiv:1601.00653 [hep-ph]].

\bibitem{nulike}
  https://nulike.hepforge.org/

\bibitem{Scott:2012mq}
  P.~Scott {\it et al.} [IceCube Collaboration],
  JCAP {\bf 1211}, 057 (2012)
  [arXiv:1207.0810 [hep-ph]].

\bibitem{Amole:2015pla}
  C.~Amole {\it et al.} [PICO Collaboration],
  Phys.\ Rev.\ D {\bf 93}, no. 5, 052014 (2016)
  doi:10.1103/PhysRevD.93.052014
  [arXiv:1510.07754 [hep-ex]].

\bibitem{Amole:2016pye}
  C.~Amole {\it et al.} [PICO Collaboration],
  Phys.\ Rev.\ D {\bf 93}, no. 6, 061101 (2016)
  doi:10.1103/PhysRevD.93.061101
  [arXiv:1601.03729 [astro-ph.CO]].

\bibitem{Savage:2008er}
  C.~Savage, G.~Gelmini, P.~Gondolo and K.~Freese,
  JCAP {\bf 0904}, 010 (2009)
  [arXiv:0808.3607 [astro-ph]];
  R.~Bernabei {\it et al.} [DAMA and LIBRA Collaborations],
  Eur.\ Phys.\ J.\ C {\bf 67}, 39 (2010)
  [arXiv:1002.1028 [astro-ph.GA]].



\bibitem{Spitz:2012gp}
  J.~Spitz,
  Phys.\ Rev.\ D {\bf 85}, 093020 (2012)
  [arXiv:1203.6050 [hep-ph]].

\bibitem{Conrad:2010eu}
  J.~M.~Conrad,
  Nucl.\ Phys.\ Proc.\ Suppl.\  {\bf 229-232}, 386 (2012)
  [arXiv:1012.4853 [hep-ex]].




\end{thebibliography}
\end{document}